\documentclass[twocolumn,showpacs,preprintnumbers,amssymb]{revtex4}

\usepackage{graphicx}
\usepackage{dcolumn}
\usepackage{bm}

\begin{document}

\preprint{report.tex}

\title{Surface term for the capillary condensation transitions in a
slit geometry}

\author{Ignacio Urrutia}
 \altaffiliation[Also at ]{Carrera del Investigador Cient\'{\i}fico
of the Consejo Nacional de Investigaciones Cient\'{\i}ficas y
T\'ecnicas, Av. Rivadavia 1917, RA--1033 Buenos Aires, Argentina}

\author{Leszek Szybisz}
 \altaffiliation[Also at ]{Carrera del Investigador Cient\'{\i}fico
of the Consejo Nacional de Investigaciones Cient\'{\i}ficas y
T\'ecnicas, Av. Rivadavia 1917, RA--1033 Buenos Aires, Argentina}

\affiliation{Laboratorio TANDAR, Departamento de F\'{\i}sica,
Comisi\'on Nacional de Energ\'{\i}a At\'omica,\\
Av. del Libertador 8250, RA-1429 Buenos Aires, Argentina}
\affiliation{Departamento de F\'{\i}sica, Facultad de
Ciencias Exactas y Naturales,\\
Universidad de Buenos Aires,
Ciudad Universitaria, RA-1428 Buenos Aires, Argentina}

\date{\today}

\begin{abstract}
It is shown that a bare simple fluid model (SFM) proposed some
years ago for studying adsorption between two semi-infinite
solid walls can be improved by modifying the surface term in
the grand potential for the film phase. Such a correction
substantially improves the agreement between the predictions
for phase transitions provided by that SFM and results obtained
from calculations carried out for $^4$He with the
density-functional method at zero temperature. The corrective
term depends on the strength of the adsorption potential and
observables of bulk helium.

\end{abstract}


\maketitle

A simple fluid model (SFM) has been proposed by Gatica {\it et
al.}~\cite{gat0} for exploring the behavior of adsorption between
two parallel walls separated by a distance $L$. In that paper,
transitions between empty (E), film (F), and capillary condensation
(CC) phases were determined. Subsequently, Calbi 
{\it et al.}~\cite{calb} have checked such a SFM by comparing its
predictions with results provided by the nonlocal
density-functional (NLDF) formulated by the Orsay-Paris (OP)
collaboration \cite{dupont90} for superfluid helium. By looking at
Figs.\ 15 and 16 in Ref.~\onlinecite{calb} one may realize that
the E {$\to$} F phase transitions given by SFM for $^4$He/Li and
$^4$He/Au systems do not reproduce quite well results obtained
from the OP-NLDF theory. In this report we suggest a correction
which substantially improves the agreement.

For the descriptions provided in Refs.~\onlinecite{gat0} and
\onlinecite{calb} it is assumed that $^4$He atoms interact with
the walls via the net potential
\begin{equation}
U_{\rm slit}(z) = V(z) + V(L-z) \:, \label{Uuu}
\end{equation}
with $V(z)$ being the standard (9,3) adsorption potential
\begin{equation}
V(z) = [4C^3_3/(27D^2)]z^{-9} - C_3 z^{-3}\:, \label{Pote}
\end{equation}
where $D$ is the well depth and $C_3$ is the strength of the
asymptotic van der Waals interaction \cite{bruch}. This
potential $V(z)$ exhibits a minimum at
\begin{equation}
z_m = [2C_3/(3D)]^{1/3} \:. \label{min}
\end{equation}
A sample of parameters $D$ and $C_3$ corresponding to the
interaction of $^4$He atoms with several substrates given in
Refs.~\onlinecite{chen} and \onlinecite{chiz} is listed in 
Table~\ref{tab:table1}. The reader may find a plot of $D$
{\rm vs} $C_3$ in Fig. 2 of Ref.~\onlinecite{chen}. It should
be mentioned that for alkali metals the adsorption potentials
of Ref.\ \onlinecite{chiz} are more attractive than the
previous ones of Ref.~\onlinecite{chen}.

\begin{table}
\caption{\label{tab:table1}Values of the well depth $D$, the
van der Waals strength $C_3$, the position of the potential's
minimum $z_m$, the reduced well depth $D^*$ and other relevant
quantities.}
\begin{ruledtabular}
\begin{tabular}{lrrcrcccc}
Surface & $D$\footnotemark[1] & $C_3$\footnotemark[1] & $z_m$
& $D^*$ & $\sigma^*$ & $\beta_{\rm PH}$ & $\beta_{\rm HO}$
& $\beta_{\rm MHO}$ \\
& [K] & [K\AA$^3$] & [\AA] &&&&& \\
\hline
Na\footnotemark[2] &  12.53 & 1197.4 & 3.99 &  8.03 & 0.2177
&      & 2.65 & 2.18 \\
Li\footnotemark[2] &  17.87 & 1422.5 & 3.76 & 10.78 & 0.2054
& 2.80 & 2.81 & 2.53 \\
H$_2$ &  28.00 &  360.0 & 2.05 &  9.20 & 0.2487 & 2.23 & 2.32 & 2.34 \\
Mg\footnotemark[2] &  35.63 & 1850.2 & 3.26 & 18.64 & 0.1856
& 3.12 & 3.11 & 3.32 \\
Ne    &  60.00 &  163.0 & 1.22 & 11.74 & 0.2664 & 2.78 & 2.16 & 2.64 \\
Al    &  60.30 & 2340.0 & 2.96 & 28.63 & 0.1708 & 3.77 & 3.38 & 4.12 \\
Cu    &  63.80 & 2610.0 & 3.01 & 30.83 & 0.1669 & 3.90 & 3.46 & 4.28 \\
Ag    &  63.80 & 2890.0 & 3.11 & 31.90 & 0.1641 & 3.95 & 3.52 & 4.35 \\
NaCl  &  69.60 & 1230.0 & 2.28 & 25.42 & 0.1879 & 3.73 & 3.07 & 3.88 \\
Ar    &  78.90 &  870.0 & 1.94 & 24.63 & 0.1970 & 3.83 & 2.93 & 3.82 \\
Au    &  92.80 & 3180.0 & 2.84 & 42.27 & 0.1566 & 4.59 & 3.69 & 5.01 \\
LiF   &  94.00 & 1080.0 & 1.97 & 29.75 & 0.1872 & 4.26 & 3.08 & 4.20 \\
Gr    & 190.00 & 2130.0 & 1.96 & 59.64 & 0.1577 & 6.63 & 3.66 & 5.95 \\
\end{tabular}
\end{ruledtabular}
\footnotemark[1]{Potential parameters $D$ and $C_3$ are taken from
Ref. \onlinecite{chen} except where noted.}
\footnotemark[2]{Potential parameters from Ref.\ \onlinecite{chiz}.}
\end{table}

According to SFM, the E {$\to$} F transition for films of
thickness $\ell$ adsorbed on both walls of the slit is governed
by the evolution of the reduced ({\it i.e.}, in units of the
surface tension $\sigma_{lv}=0.272$~K\AA$^{-2}$ \cite{szyb00})
grand potential per unit area given by Eq.\ (8) of
Ref.~\onlinecite{gat0}
\begin{eqnarray}
\Omega^*_{\rm F} = \frac{\Omega_{\rm F}}{\sigma_{lv}}&=&4
- D^*\,[g\{1+x\} + g\{L^*-1\}
\nonumber\\
&&~- g\{L^*-1-x\}] + 2\,x\,\Delta \:. \label{omegas0}
\end{eqnarray}
Here the first term is the surface energy, the second is due to
the helium-walls interaction, and the last one measures the
departure from the saturated vapor pressure (SVP) condition.
The reduced well depth and gap are
\begin{equation}
D^* = 2\,\rho _0\,D\,z_m/\sigma_{lv} \:, \label{Dee}
\end{equation}
\begin{equation}
\Delta = (\mu_0-\mu)\,\rho_0\,z_m/\sigma_{lv} \:, \label{Del}
\end{equation}
where $\rho_0=0.021836$~\AA$^{-3}$ is the saturation equilibrium
density and $\mu_0=-7.15$~K the corresponding chemical potential
\cite{szyb00}. Furthermore,
\begin{equation}
g\{y\} = \int^y_1 dz^*\,[\frac{-U_{\rm slit}(z_m\,z^*)}
{D}] = \frac{11}{16} - \frac{3}{4}\frac{1}{y^2} + \frac{1}{16}
\frac{1}{y^8} \:, \label{gee}
\end{equation}
here all the reduced distances are given in terms of $z_m$
\begin{equation}
z^* = z/z_m,~~~x = \ell/z_m \:,~~~and~~~L^* = L/z_m \:,
\label{Ly}
\end{equation}
where $x$ is the film thickness that minimizes $\Omega^*_{\rm F}$.

\begin{figure}
\includegraphics[width=6.0cm, angle=-90]{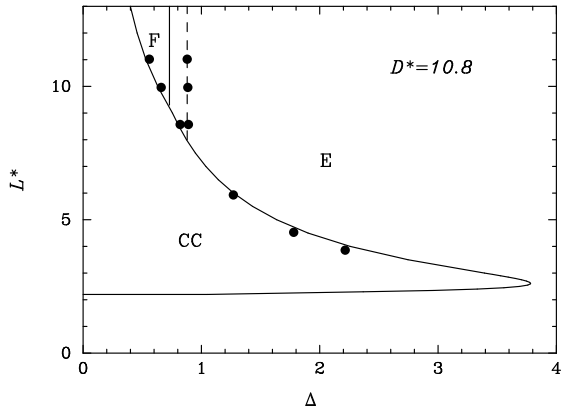}
\caption{\label{fig:Li}Reduced phase diagram for $^4$He confined
by two planar walls of Li ($D^*=10.78$, $z_m=3.76$ \AA). Solid
curves are SFM predictions and full circles OP-NLDF results,
both of them calculated in the present work (see text). The
dashed curve shows how much the correction introduced in the
present work improves the prediction of the SFM for the E
{$\to$} F phase transition.}
\end{figure}

\begin{figure}
\centering\includegraphics[width=6cm, angle=-90]{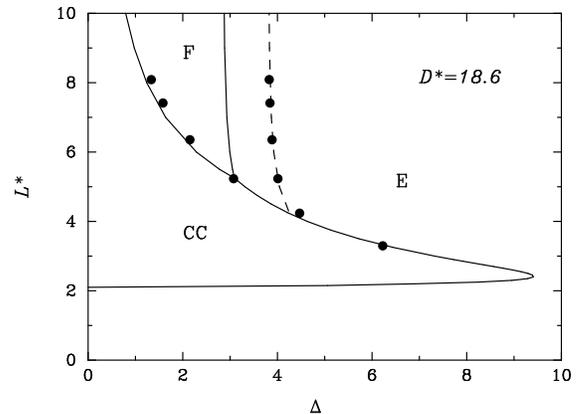}
\caption{\label{fig:Mg}Same as Fig.\ \protect\ref{fig:Li} but for
$^4$He confined by Mg ($D^*=18.64$, $z_m=3.26$ \AA).}
\end{figure}

\begin{figure}
\centering\includegraphics[width=6cm, angle=-90]{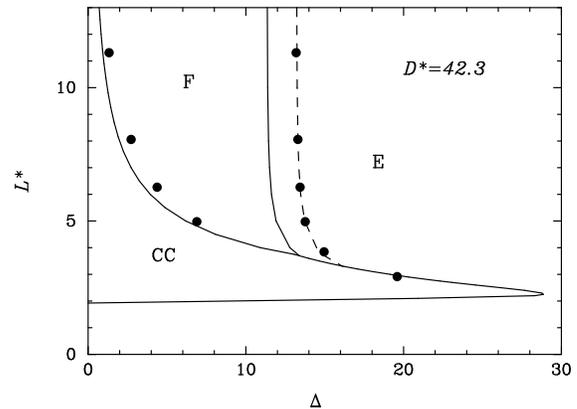}
\caption{\label{fig:Au}Same as Fig.\ \protect\ref{fig:Li} but for
$^4$He confined by Au ($D^*=42.27$, $z_m=2.84$ \AA).}
 \end{figure}

Figure \ref{fig:Li} shows that according to present calculations
for the case of $^4$He confined into slits of Li the reduced
grand potential $\Omega^*_{\rm F}$ given by the SFM becomes
negative for $L^* > 9$ at $\Delta_{\rm SFM}=0.73$, while the
OP-NLDF predicts that the E {$\to$} F phase transition occurs at
$\Delta_{\rm DF}=0.85$. Similar effects are also obtained for
that substrates listed in Table~\ref{tab:table1} which are
strong enough to produce a stable monolayer film. In order to
illustrate this feature results for $^4$He/Mg and $^4$He/Au are
plotted in Figs. \ref{fig:Mg} and \ref{fig:Au}. Let us recall
that a slit of Na is too weak to produce a stable monolayer
structure.

In Figs. \ref{fig:Li}-\ref{fig:Au} besides the E {$\to$} F phase
transition there are also indicated the E {$\to$} CC and F {$\to$}
CC ones. These phase transitions are determined from comparisons
with the corresponding grand potential per unit area obtained from
Eq.\ (5) of Ref.~\onlinecite{gat0}, i.e.,
\begin{equation}
\Omega^*_{\rm CC} = \frac{\Omega_{\rm CC}}{\sigma_{lv}} = 2
- D^*\,g\{L^*-1\} + (L^*-2)\,\Delta \:. \label{omegaCC}
\end{equation}

In this work we shall demonstrate that the prediction given by the
bare SFM is improved by introducing a correction to the surface
term in the grand potential $\Omega^*_{\rm F}$ given by Eq.\ 
(\ref{omegas0}). An important shortcoming is that this term does
not vanish in the limit of a zero-thickness film (i.e. for $x =
\ell/z_m \to 0$). In order to eliminate this failure one may follow
the idea adopted by Cheng {\it et al.}\cite{chen0} in writing their
Eq.\ (2.4). So that, we shall assume that the surface contribution
grows exponentially from zero at $\ell=0$ to the bare value $4$ over
a characteristic length $\zeta$
\begin{equation}
\Omega^*_{\rm F}({\rm surf}) = 4\,[ 1 - \exp(-\ell/\zeta) ] = 4\,
[ 1 - \exp\left(-\beta\,x\right) ] \:, \label{omegas1}
\end{equation}
with
\begin{equation}
\beta = z_m/\zeta = 1/\zeta^* \:. \label{beta}
\end{equation}

The parameter $\beta$ was determined for all the systems quoted
in Table~\ref{tab:table1} by requiring that OP-NLDF results be
reproduced by SFM. This means, we impose that the E {$\to$} F
transition must occur at $\Delta_{\rm DF}$ for both models. These
phenomenological values denoted $\beta_{\rm PH}$ are also included
in Table~\ref{tab:table1}. The changes of the energetics due to
the cut-off factor introduced in Eq.\ (\ref{omegas1}) only become
important for rather thin films.

It should be also noted that, if a similar correction is included
in the grand potential for the CC phase the corrected curves for
the E {$\to$} CC and F {$\to$} CC transitions do not differ much
from that displayed in Figs. \ref{fig:Li}-\ref{fig:Au}.

The following lines are devoted to propose a possible origin for
the cut-off factor $\beta=z_m/\zeta$. As a first step, it is
plausible to assume that the characteristic length $\zeta$ be
related to the width of the ground-state (g.s.) wavefunction
(submonolayer solution) of a simple harmonic-oscillator
approximation to the adsorption potential in the neighborhood
of each wall
\begin{equation}
V_{\rm app}(z) = - D + V_{\rm HO}(z) = - D
+ \frac{1}{2}\,k\,(z-z_m)^2 \:, \label{ho}
\end{equation}
with the force constant
\begin{equation}
k = \biggr[\frac{d^2 V}{d z^2}\biggr]_{z=z_m}
= 18\,C_3 \left(\frac{3\,D}{2\,C_3}\right)^{5/3}
= 27\,\frac{D}{z^2_m} \:, \label{elastic}
\end{equation}
see comment quoted as citation $44$ in Ref.\ \onlinecite{chen0}.
The dimensionless force constant becomes
\begin{equation}
\kappa = k\,z^2_m/D = 27 \:. \label{elastics}
\end{equation}
So, the reduced version of $V_{\rm app}(z)$ reads
\begin{equation}
V^*_{\rm app}(z^*) = \frac{V_{\rm app}(z_m\,z^*)}{D} = -1 +
\frac{27}{2}\,(z^*-1)^2 \:, \label{ho1}
\end{equation}
and the reduced hamiltonian may be conveniently written as
\begin{equation}
H^*_{\rm HO} = H^*_{\rm app} + 1 = - \frac{\hbar^2}{2\,
m_{\rm eff}}\,\frac{d^2}{{dz^*}^2} + \frac{27}{2}\,(z^*-1)^2
\:, \label{ho2}
\end{equation}
with an effective helium mass
\begin{equation}
m_{\rm eff} = m\,D\,z^2_m \:. \label{mass}
\end{equation}
The g.s. energy of $H^*_{\rm app}$ is
\begin{equation}
E^*_{\rm app} = -1 + E^*_{\rm HO} = -1 + \frac{1}{2}
\frac{\hbar}{D}\sqrt{\frac{k}{m}} = -1 + \frac{\sqrt{27\,E^*_k}}{2}
\:, \label{Eho}
\end{equation}
where $E^*_k=\hbar^2/m_{\rm eff}$ is a dimensionless kinetic
energy factor. Let us mention that as quoted in Ref.\
\onlinecite{vidali}, for graphite (Gr) the semi-empirical He
potential of Carlos and Cole \cite{caco}, based on He scattering
data, has $\kappa = 26$.

\begin{figure}
\centering\includegraphics[width=5.5cm, angle=-90]{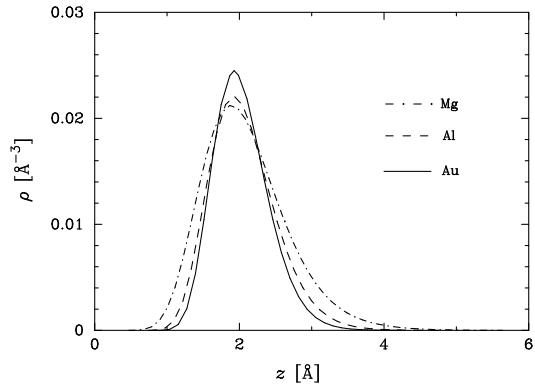}
\caption{\label{fig:rhos}Comparison of monolayer $^4$He systems
adsorbed on the left wall of broad slits of Mg, Al, and Au.
These films were determined at the E $\to$ F phase transition,
i.e., they correspond to the lowest stable coverage in each
case.}
\end{figure}

The probability density of the g.s. solution of Eq.\ ($\ref{ho2}$)
is
\begin{equation}
\mid \psi_0(z^*) \mid^2 = \frac{1}{\sqrt{2\,\pi}\,\sigma^*}
\,e^{-\frac{1}{2}[(z^*-1)/\sigma^*]^2} \:, \label{funk}
\end{equation}
with
\begin{equation}
\sigma^* = \frac{\sigma}{z_m}
= \frac{1}{z_m} \left(\frac{\hbar^2}{m\,k}\right)^{1/4}
= \left(\frac{E^*_k}{27}\right)^{1/4} \:. \label{width0}
\end{equation}
Figure \ref{fig:rhos} shows examples of density profiles yielded
by the DF theory for the lowest stable coverages. Let us recall
that in the literature the surface thickness is defined as the
distance over which the density falls off from the 90\% to 10\% of
the saturation value \cite{sur2}. In the case of the g.s.
wavefunction given by Eq. (\ref{funk}) these quota lead to
\begin{equation}
{\cal R}(\zeta^*_i) = \frac{\mid \psi_0(1+\zeta^*_i) \mid^2}{\mid
\psi_0(1) \mid^2 } = e^{-\frac{1}{2}[\zeta^*_i/\sigma^*]^2}
= 0.90 \:, \label{ene1}
\end{equation}
\begin{equation}
{\cal R}(\zeta^*_f) = \frac{\mid \psi_0(1+\zeta^*_f) \mid^2}{\mid
\psi_0(1) \mid^2 } = e^{-\frac{1}{2}[\zeta^*_f/\sigma^*]^2}
= 0.10 \:. \label{ene2}
\end{equation}
From here one gets
\begin{eqnarray}
\zeta^* = \zeta^*_f - \zeta^*_i&=& \frac{4\,\ln3}{\zeta^*_f
+ \zeta^*_i}(\sigma^*)^2
= \frac{2\sqrt{2} \ln3}{\sqrt{\ln 10}+\sqrt{\ln(10/9)}}
\sigma^* \nonumber\\
&=&1.69\,\sigma^* \simeq \sqrt{3}\,\sigma^* \:. \label{zeda1}
\end{eqnarray}
It is interesting to notice that by assuming $\zeta^* \simeq
\sqrt{3}\,\sigma^*$ the integral
\begin{equation}
{\cal I}(\zeta^*) = \frac{1}{\sqrt{2\,\pi}\,\sigma^{*}}
\int^{1+\zeta^*}_{1-\zeta^*} dz^* \,
e^{-\frac{1}{2}[(z^*-1)/\sigma^*]^2} \simeq 0.90 \:, \label{cee}
\end{equation}
amounts about $90\%$ of the total norm.

Turning to the cut-off parameter $\beta$, the expression derived
from a pure harmonic oscillator analysis becomes
\begin{equation}
\beta_{\rm HO} = \frac{1}{\zeta^*}
= \frac{1}{\sqrt{3}}\,\left(\frac{27}{E^*_k}\right)^{1/4}
= \left(\frac{3\,m_{\rm eff}}{\hbar^2}\right)^{1/4} \:.
\label{beta0}
\end{equation}
The estimations provided by this formula are listed in
Table~\ref{tab:table1}. From a glance at that table, one can
realize that there is a very good agreement for the most weakly
bounded monolayer systems, i.e., that confined by walls of H$_2$,
Li and Mg. It is worthwhile to mention that the values of $\beta$
for $^4$He/Li and $^4$He/Mg are in the parameter region utilized
by Cheng {\it et al.} \cite{chen0} for showing how SFM results
approach those given by a full NLDF theory in the case of
adsorption on semi-infinite surfaces with planar geometry. In
particular, $\beta=3$ is just the value chosen in that paper to
illustrate the behavior.

\begin{figure}
\centering\includegraphics[width=6cm, angle=-90]{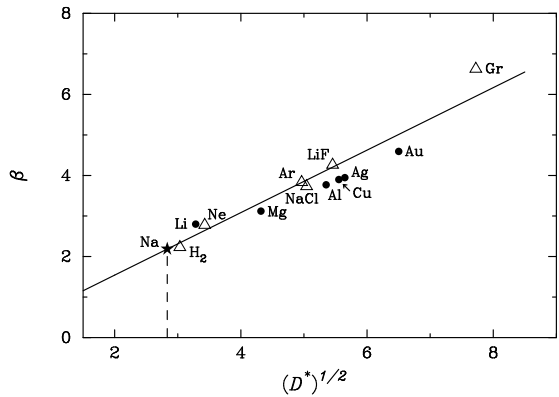}
\caption{\label{fig:betag}Parameter $\beta_{\rm PH}$ obtained
from equating SFM and DF results for $^4$He films as a function
of $\sqrt{D^*}$. Full circles are metallic substrates, while
open triangles stand for other kind of materials. The star is
the $\beta_{\rm MHO}$ value for a slit of Na. The solid straight
line is given by Eq. (\protect\ref{beta2}).}
\end{figure}

The values of $\beta_{\rm HO}$ fall short if the helium atoms are
exposed to stronger attractions. Hence, one could argue that a
correction proportional to $D$ should be introduced if the
harmonic oscillator approach for the potential $V(z)$ of Eq.\
(\ref{Pote}) yields a g.s. oscillator energy much smaller than
the well depth, i.e., when in Eq.\ (\ref{Eho}) holds $\frac{1}{2}
\,\frac{\hbar}{D}\,\sqrt{\frac{k}{m}} \ll 1$. In searching
for such a dependence, it was realized that the values of
$\beta_{\rm PH}$ become aligned when plotted as a function of
$\sqrt{D^*}$. This behavior may be observed in Fig.\
\ref{fig:betag}, where Na indicates the threshold obtained for
the formation of stable monolayer films. Similar results are
obtained by using the more recent NLDF of Ref.\
\onlinecite{dalfo95}.

Therefore, as a next step the relative influence of effects due
to the strength of the adsorption potential and to the resistance
of helium atoms to be compressed was analyzed. It was found that
if one modifies $\beta_{\rm HO}$ by including a factor which
takes into account the ratio between the well depth $D$ and the
compressibility at saturation density $1/\rho_0\kappa_v=27.2$~K
\cite{szyb00} in such a way that the total effect be
proportional to $\sqrt{D^*}$
\begin{eqnarray}
\beta_{\rm MHO}&=&\beta_{\rm HO}
\left(\frac{D}{1/\rho_0\kappa_v}\right)^{1/4} \nonumber\\
&=&\left(\frac{3\,m\,D\,z^2_m}{\hbar^2}\right)^{1/4}
\times \left(\frac{D}{1/\rho_0\kappa_v}\right)^{1/4} \nonumber\\
&=&\left(\frac{3}{4}\frac{m}{\hbar^2}\frac{\sigma^2_{lv}}
{\rho_0}\,\kappa_v\right)^{1/4} \sqrt{D^*} \;, \label{beta2}
\end{eqnarray}
one gets a good overall agreement as shown in Fig.\
\ref{fig:betag}. It is worthy of notice that the slope of this
simple expression is written in terms of physical observables
of bulk helium.

Moreover, it is important to emphasize that Fig.\ \ref{fig:betag}
collects data on substrates from H$_2$ and Li, which are barely
able to support monolayer helium films, up to Gr (graphite), which
is the strongest attractor for helium atoms \cite{bruch}. Equation
(\ref{beta2}) can be considered as a {\it nearly} universal
property since the straight line for $\beta_{\rm MHO}$ matches
well the data of $\beta_{\rm PH}$ for all the substrates included
in the present Table~\ref{tab:table1}, i.e., for those plotted in
Fig. 2 of Ref.\ \onlinecite{chen} provided they ly above the dashed
curve traced there. Notice that in that figure the points
determined by $D$ and $C_3$ are spread over the whole drawing.

In conclusion, one can state that a simple semi-phenomenological
correction in the surface term of the SFM given by Eq.\
(\ref{omegas0}) yields an important improvement of the prediction
for the E $\to$ F phase transition. The corrective term proposed
in the present report contains a decaying exponential built by
following the simple model formulated by Cheng {\it et al.}
\cite{chen0}. The cutoff parameter can be interpreted by taking
into account properties of the harmonic g.s. wavefunction, the
well depth of the adsorption potential and the compressibility
of bulk helium.

\begin{acknowledgments}
This work was supported in part by the Ministry of Culture and
Education of Argentina through Grants CONICET PIP No. 5138/05
and UBACyT No. X298.
\end{acknowledgments}

\end{document}